\input{aipcheck}

\documentclass[
    ,final            
  ]
  {aipproc}

\layoutstyle{8x11double}

\begin{document}

\title{Orbifold SUSY GUT from the Heterotic String}


\author{Bumseok Kyae}{
  address={Department of Physics and Astronomy and Center for Theoretical Physics,
\\ Seoul National University, Seoul 151-747, Korea
}
}


\def\lsim{\lower.7ex\hbox{$\;\stackrel{\textstyle<}{\sim}\;$}}
\def\lsl{ l \hspace{-0.45 em}/}
\def\ksl{ k \hspace{-0.45 em}/}
\def\qsl{ q \hspace{-0.45 em}/}
\def\psl{ p \hspace{-0.45 em}/}
\def\ppsl{ p' \hspace{-0.70 em}/}
\def\dsl{ \partial \hspace{-0.5 em}/}
\def\Dsl{ D \hspace{-0.55 em}/}
\def\N{$\cal N$}
\def\tphi{\tilde\phi}

\def\Qem{{$Q_{\rm em}$}}
  \def\SMSSM{${\cal S}$MSSM\ }
\def\CPT{${\cal CPT}$\ }

 \def\N{{$\cal N$}}
 \def\Z{{\bf Z}}
 \def\MG{{$M_{\rm GUT}$}}

\def\EE{E$_8\times$E$_8^\prime$}

 \def\N{{$\cal N$}}
 \def\Z{{\bf Z}}
 \def\MG{{$M_{\rm GUT}$}}

\def\EE{E$_8\times$E$_8^\prime$}
\def\Eo{E$_8$}
\def\Eh{E$_8'$}
\def\MGUT{$M_{\rm GUT}$}
\def\ie{{\it i.e.}\ }

\def\fourb{\overline{\bf 4}}
\def\four{{\bf 4}}
\def\two{{\bf 2}}
\def\fiveb{\overline{\bf 5}}
\def\five{{\bf 5}}
\def\tenb{\overline{\bf 10}}
\def\ten{{\bf 10}}
\def\one{{\bf 1}}
\def\six{{\bf 6}}
\def\threeb{\overline{\bf 3}}
\def\three{{\bf 3}}


\begin{abstract}
From the string partition function, we discuss the mass-shell and
GSO projection conditions valid for Kaluza-Klein (KK) as well as
massless states in the heterotic string theory compactified on a
nonprime orbifold. Using the obtained conditions we construct a 4D
string standard model, which is embedded in a 6D SUSY GUT by
including KK states above the compactification scale. We discuss
the stringy threshold corrections to gauge couplings, including
the Wilson line effects.
%
%
%
%
%
%
%
\end{abstract}

\maketitle



\noindent {\bf Introduction.}\quad An indirect but the most
convincing evidence for the existence of the supersymmetric grand
unified theory (SUSY GUT) would be found from the notice that the
three gauge couplings in the minimal supersymmetric standard model
(MSSM), $\{g_3, g_2, \sqrt{5/3}g_Y\}$
%
%
are unified at the $10^{16}$ GeV scale. Here the non-trivial
normalization factor ``$\sqrt{5/3}$'' of $g_Y$, which is essential
for the unification, is best explained in the framework of GUT. It
seems to imply the real presence of GUT at the $10^{16}$ GeV
scale.

A realization of the GUT idea, however, would be quite complicated
in the 4 dimensional (4D) spacetime:  Even spontaneous symmetry
breaking mechanism of GUT into the standard model (SM) gauge
symmetry $G_{SM}$ without fine-tuning and without leaving unwanted
pseudo-Goldstones and triplet Higgs is not simple. Accordingly,
even the gauge coupling unification consistent with the low energy
value of sin$^2\theta_W$ ($\approx 0.23$) in GUT, which provided a
strong evidence supporting GUT, is non-trivial.

In higher dimensional spacetime, however, SUSY GUT idea can be
simply realized, if the compactification mechanism of the extra
dimensions is associated with GUT and SUSY breakings. For
instance, in 5D orbifold compactification with $S^1/(Z_2\times
Z_2')$~\cite{5dSUSYGUT}, the boundary conditions consistent with
$Z_2$ parities lead to ${\cal N}=2$ SUSY breaking (in terms of 4D
SUSY) into ${\cal N}=1$. The other $Z_2'$, which is originated
from the translational symmetry of $S^1$, can be optionally
utilized to break the gauge symmetry $G_{\rm GUT}$. As a result,
below the compactification scale ($\approx 1/R$) we have ${\cal
N}=1$ MSSM, while above the $1/R$ scale the original $G_{\rm GUT}$
as well as ${\cal N}=2$ SUSY is effectively restored, because of
the presence of Kaluza-Klein (KK) modes. Unlike in the 4D GUTs,
the notorious doublet/triplet splitting problem is also very
easily resolved (just by boundary conditions). Thus, this model
maintains the nice features of the GUT idea in a simple way.
Unfortunately, however, 5D orbifold SUSY GUT is not renormalizable
and so quantum mechanically not predictable. It needs to be
supported by the fundamental theory such as the string theory.

\noindent {\bf Nonprime Orbifold.}\quad By employing the {\it
nonprime} orbifold compactification of the heterotic string
theory, we can obtain 5D or 6D orbifold SUSY GUTs. In the large
limit of the extra 2D (1D), a 4D theory would become an effective
6D (5D) string theory with ${\cal N}=2$ or 4 SUSY. In the orbifold
compactification, the ${\cal N}=2$ SUSY  requires the presence of
a sub-lattice invariant under a given twist action, i.e. a torus.
Hence, there is no twisting for strings on it. Background moduli
such as radius (or metric) can be encoded only in the zero modes'
momenta of untwisted bosonic strings~\cite{Dixon:1990pc}, which
are interpreted as KK modes in the 4D spacetime. Since twisted
strings are stuck to orbifold fixed points, they cannot
accommodate moduli. In order to introduce a modulus such as
arbitrarily large radius and to discuss the relevant KK
excitations, therefore, we need to employ an orbifold
compactification providing an invariant sub-lattice.

In nonprime orbifold compactifications, some higher twist sectors
turn out to behave like in the invariant sub-lattice preserving
${\cal N}=2$ SUSY. KK excited states can arise from such higher
twist sectors in nonprime orbifold
compactifications~\cite{Dixon:1990pc}.  Moreover, orbifold {\it
string} compactifications must break a gauge symmetry as well as
SUSY, because of the modular invariance of the string theory.
Accordingly, the gauge symmetry could be enhanced by including KK
massive states above the $1/R$ scale. Hence, it is possible to
construct a higher dimensional SUSY GUT with the help of KK
states~\cite{Z12I,SMZ12I,Lebedev}. We are particularly interested
in the nonprime orbifold $\Z_{12-I}$ which has some interesting
features \cite{KimKyaeKK08,Z12I,SMZ12I}. This method can be easily
extended to the other nonprime orbifold compactifications.

The $\Z_{12-I}$ orbifold is defined with the twist vector
$\phi=\left(\frac{5}{12},\frac{4}{12},\frac{1}{12}\right)$, acting
on the complexified three dimensional string:
$X^A_{L,R}(2\pi)=e^{+2\pi i\phi_A}X_{L,R}^A(0)$, $A=1,2,3$.
Similarly, $X_{L,R}^{A*}$ acquires $e^{-2\pi i\phi_A}$. This
orbifold can be the ${\rm SO(8)\times SU(3)}$ or ${\rm F_4\times
SU(3)}$ lattice, where the SU(3) lattice corresponds to the second
complex plane. We will assume that the size $R$ of the second
torus is relatively larger than the sizes of the first and the
second tori.

The modular invariance of the orbifolded string theory requires to
consider all the possible sectors with the boundary conditions,
$0\times\phi,~1\times\phi,~2\times\phi,\cdots, 11\times\phi$,
which are called ``$U$'', ``$T_1$,'' ``$T_2$,''$\cdots,$
``$T_{11}$'' sectors, respectively. The sector of $12\times\phi$
is identified with the $U$ sector.
In the $T_3$, $T_6$, and $T_9$ sectors of $\Z_{12-I}$, where the
boundary conditions are given by $3\phi$, $6\phi$, and $9\phi$,
respectively, the second (complexified) sub-lattice remains
untwisted.  Namely, the second sub-lattice is just an ordinary
torus in the $T_3$, $T_6$, and $T_9$ sectors. The KK massive
states can arise only from the sectors associated with invariant
sub-lattices, and they indeed respect \N=2
SUSY~\cite{Dixon:1990pc,KimKyaeKK08}.

\noindent {\bf Partition Function.}\quad Many important physical
information such as the mass-shell condition, GSO projection, and
so on, can be extracted from the one loop partition function. The
one-loop amplitude by closed strings has the topology of a torus.
A world-sheet torus can be parametrized by $\sigma_1+\tau\sigma_2$
($0\leq\sigma_1,~\sigma_2\leq 2\pi$), where $\tau$ ($\equiv
\tau_1+i\tau_2$) is the modular parameter. On the  world-sheet
torus, we have two boundary conditions, $[g^k,~g^l]$ (or simply
$[k,l]$), where $g^k$ ($g^l$) implies the action of the order $k$
($l$) twisted string boundary conditions in the $\sigma_1$
($\sigma_2$) direction. In the $[k,l]$ sectors with $k~and~l=
0,3,6,9$, $X_{L,R}^i$ in the longitudinal directions of the second
complex plane ($i=3,4$)  are untwisted, and so KK massive states
can arise from there.  On the other hand, in the $[k,l]$ sectors
with $k~or~l\neq 0,3,6,9$, where only ${\cal N}=1$ SUSY is
preserved, there do not appear KK excited states.

The full partition function, which is given by the summation of
the $[k,l]$ sectors' contributions, should be invariant under the
modular transformations, i.e. ``S'' ($\tau\rightarrow -1/\tau$)
and ``T'' ($\tau\rightarrow\tau+1$) transformations. Under the S
(T) transformation, a $[k,l]$ sector is cast to the $[l,-k]$
($[k,l+k]$) sector.  For $\tau\rightarrow -1/\tau$ and
$\tau\rightarrow\tau+1$, therefore, the $[k,l]$ sectors with
$k,l=0,3,6,9$  are interchanged each other only inside $\{[k,l]$
sectors ; $k,l=0,3,6,9\}$, decoupled from the other sectors with
$k,l\neq 0,3,6,9$.  The partition function invariant under the S
and T transformations in $U$, $T_3$, $T_6$, $T_9$ takes typically
the following form~\cite{KimKyaeKK08}:
\begin{eqnarray} \label{full}
&&\quad\quad \frac{1}{\lambda^2}\Bigg[\sum_{\vec{\mu}\in
\Lambda^*_2,\vec{\zeta}\in\Lambda_2}
\frac{q^{(\vec{L}+\vec{\zeta})^2/2}}{\left[\eta(\tau) \right]^2}
~\frac{\bar{q}^{(\vec{L}-\vec{\zeta})^2/2}}
{\left[\bar{\eta}(\bar{\tau})\right]^2}\Bigg]
\times \\
&&\quad \Bigg[\sum_{P\in\Lambda_{16},s\in\Lambda_8}
\frac{q^{(L^I)^2/2}}{\left[\eta(\tau)\right]^{16}}
~\frac{\bar{q}^{(\tilde{L})^2/2}}{\left[\bar{\eta}
(\bar{\tau})\right]^4} \times e^{2\pi il\Theta_k}\Bigg]
\big|\widehat{Z}^X_{[k,l]}\big|^2 . \nonumber
\end{eqnarray}
Here we neglect the spin structure for simplicity. $\vec{L}$,
$L^I$, and $\Theta_k$ are given by
\begin{eqnarray}
\vec{L}=\sum_{a,b=3,4}\bigg[ \frac{m_a\vec{e}^{*a}}{2\lambda}
-\left(P^I+\frac{kV^I}{2}+\frac{\zeta^bW^I_b}{2}\right)
\frac{W_a^I}{2}~\vec{e}^{*a} \bigg] ~,  \label{L^i}
\end{eqnarray}
\begin{eqnarray}
L^I=P^I+kV^I+\zeta^aW^I_a ~,\quad\quad \tilde{L}=s+k\phi ~,
\label{Rmom}
\end{eqnarray}
\begin{eqnarray}
\label{theta}
&&l\times\Theta_k=l\times\left[\bigg(P^I+\frac{k}{2}V^I
+\frac{\zeta^a}{2}W^I_a\bigg)V^I-\bigg(s+\frac{k}{2}\phi\bigg)
\phi\right]
\nonumber \\
&&\quad\quad\quad +\frac{\sigma}{\lambda}m_a\zeta^a ~.
\end{eqnarray}
In Eq.~(\ref{full}) $\eta$ denotes the Dedekind (``eta'') function
and $q\equiv e^{2\pi i\tau}$. The ``bar''s means the complex
conjugates. $\zeta^a$ is a proper integer associated with winding.
The dual vector $\vec{\mu}$ ($\equiv m_a\vec{e}^{*a}$) corresponds
to the KK momentum. $P$ and $s$ indicate the \EE\ and the SO(8)
weight vectors, respectively. $V^I$ and $W^I_a$ ($I=10, 11,\cdots,
15$, $a=3, 4$) stand for the shift vector and Wilson line. They
shift \EE\ lattice vectors.
In ${\bf Z}_{12-I}$, two identical order three Wilson lines can be
introduced, $W^I_3=W^I_4\equiv W^I$, and $W_1=W_2=W_5=W_6=0$. For
consistency, $12\times V^I$ and $3\times W^I_a$ should be \EE\
weight vectors, and $V^I$ and $W^I_a$ should satisfy the modular
invariance conditions:
\begin{eqnarray}
&&12(V^2-\phi^2) = {\rm even ~integer} ~,
\label{v^2} \\
&&12 V\cdot W = {\rm even ~integer} ~,
\label{vw} \\
&&12 W^2 = {\rm even ~integer} ~. \label{w^2}
\end{eqnarray}
In Eq. (\ref{full}), $\widehat{Z}^X_{[k,l]}$ denotes the
contributions of $X_{L,R}^{1,2,5,6}$ to the partition function. It
does not contain KK states associated with internal four
dimensional radii of $(x^1,x^2;x^5,x^6)$, but provides a part of
the world sheet vacuum energy. For its complete form and
$\lambda$, $\sigma$ in Eqs.~(\ref{L^i}) and (\ref{theta}), see
Ref.~\cite{KimKyaeKK08}. The large radius ($R$) dependence of the
spectrum can appear from Eqs. (\ref{full}) and (\ref{L^i}),
through the redefinition of the SU(3) basis and their dual vectors
$\vec{e}_{a}\to (R/\sqrt{\alpha'})\vec{e}_{a}$ and $\vec{e}^{*a}
\to (R^{-1}\sqrt{\alpha'})\vec{e}^{*a}$.


\noindent {\bf Massless/KK states.}\quad The powers of $q$,
$\bar{q}$ in the partition function read the mass-shell formulae.
The KK mass tower is derived as
\begin{eqnarray} \label{kkmass}
M_{\rm KK}^2 =\sum_{m_a,m_b}\frac{g^{ab}}{2R^2} \left[m_a-
P^IW^I_a\right] \left[m_b- P^IW^I_b\right] ,
\end{eqnarray}
which is of order $1/R^2$ ($\ll 1/\alpha'$) because of the inverse
metric of SU(3) lattice, $g^{ab}$. For massless states in the $U$,
$T_3$, $T_6$, $T_9$ sectors, $P^IW^I_b={\rm integer}$ should be
guaranteed.

Both massless and KK massive states should satisfy
\begin{eqnarray}
\frac{(L^I)^2}{2} +\sum_{i=j,\bar{j}}N^L_{i}\tilde{\phi}_i
-1+\frac{c_k}{2} =\frac{(\tilde{L})^2}{2}
-\frac{1}{2}+\frac{c_k}{2} =0 ~, \nonumber
\end{eqnarray}
where $i$ runs over $\{1,2,3,\bar{1},\bar{2},\bar{3}\}$, and
$\tilde{\phi}_j\equiv k\phi_j$ mod Z such that
$0<\tilde{\phi}_j\leq 1$, and $\tilde{\phi}_{\bar{j}}\equiv
-k\phi_j$ mod Z such that $0<\tilde{\phi}_{\bar{j}}\leq 1$.
$N^L_i$ indicates the oscillator numbers ($0,1,2,\cdots$).  For
$c_k$ in the world sheet vacuum energy, see
Ref.~\cite{KimKyaeKK08,Z12I}.

The generalized GSO projector is read from the coefficient of
$q^{\alpha'M_L^2/4}\bar{q}^{\alpha'M_R^2/4}$ in the partition
function:
\begin{eqnarray} \label{GSO}
{\cal P}_k=\frac{1}{N}\sum_{l}\tilde{\chi}(\theta^k,\theta^l)
e^{2\pi i l\widetilde{\Theta}_k} ~,
\end{eqnarray}
where the complete form of $\widetilde{\Theta}_k$ and the
degeneracy factor $\tilde{\chi}(\theta^k,\theta^l)$ are presented
in Ref.~\cite{KimKyaeKK08}. $N=12$ for the massless states.
However, $N=4$ for the KK massive states, because KK massive
states appear only in the $U$, $T_3$, $T_6$, and $T_9$ sectors.
Hence, $l$ in Eq.~(\ref{GSO}) runs $0,3,6,9$ for KK states,
whereas $l=0,1,2,\cdots, 11$ for massless states.


\noindent {\bf Threshold Correction.}\quad The general expression
for the moduli dependent stringy threshold correction to the gauge
couplings $\Delta_i$ can be obtained as~\cite{Dixon:1990pc}
\begin{eqnarray}
\Delta_i=\frac{|G'|}{|G|}\cdot b_i^{N=2}
\int_\Gamma\frac{d^2\tau}{\tau_2}\left(\hat{Z}_{\rm
torus}(\tau,\bar{\tau})-1\right) ,\label{threshold6D}
\end{eqnarray}
where $b_i^{N=2}$ denotes the beta function coefficient of ${\cal
N}=2$ SUSY sector by the KK modes. Since $G={\bf Z}_4\times {\bf
Z}_3$ and $G'={\bf Z}_3$ in our case,
$\frac{|G'|}{|G|}=\frac{1}{4}$. The $\hat{Z}_{\rm
torus}(\tau,\bar{\tau})$
($\equiv\sum_{\vec{P}_L,\vec{P}_R}q^{P_L^2/2}\bar{q}^{P_R^2/2}$)
is given by
\begin{eqnarray} \label{Ztorus0}
\hat{Z}_{\rm torus}(\tau,\bar{\tau})
=\sum_{\vec{\mu},\vec{\zeta}}\frac{1}{\lambda^2}~
q^{(\vec{L}+\vec{\zeta})^2/2} \bar{q}^{(\vec{L}-\vec{\zeta})^2/2},
\end{eqnarray}
where $\vec{L}$ is given in Eq.~(\ref{L^i}).

Thus, if a gauge group ${G}$ is broken to  $\tilde{H}$ by the
Wilson lines and further broken to ${H}$ by orbifolding, the
renormalized gauge couplings of ${H}$ at low energies is
\begin{eqnarray}
\frac{16\pi^2}{g^2_{H}(\mu)} &=&\frac{16\pi^2}{g^2_*}+b_{H}^0~{\rm
log}\frac{M_*^2}{\mu^2}-\frac{b_{\tilde{H}}}{4}
\left[{\rm log}\frac{R^2}{\alpha'}+1.89\right] \nonumber \\
&&+\frac{b_{\tilde{H}+G/\tilde{H}}}{4} \left[\frac{2\pi
R^2}{\sqrt{3}\alpha'}-0.30\right] . \label{gaugerun}
\end{eqnarray}
$b_{H}^0$ is the beta function coefficient contributed by ${\cal
N}=1$ SUSY sector states projected by $P^I W^I={\rm integer}$.
$b_{\tilde{H}}$ ($b_{G/\tilde{H}}$) is by the ${\cal N}=2$ SUSY
sector states projected by $P^IW^I={\rm integer}$ ($P^IW^I\neq
{\rm integer}$).

The $R^2$ coefficient is contributed by all the states charged
under ${G}$. Hence, the difference ${g_{Hi}}^{-2} -{g_{Hj}}^{-2}$
does not get an $R^2$ dependent piece if the gauge groups are
unified to ${G}$ above the $1/R$. However, the logarithmic
contribution is still present. The absolute values of the constant
and the (quadratically divergent) $R^2$ term are reliable since
their calculations are based on the fundamental theory.


\noindent {\bf Model/Conclusions.}\quad Let us propose a string
model for MSSM~\cite{SMZ12I}. It turns out to be embedded in a 6D
SU(8) GUT above the $1/R$ ($\sim M_{\rm GUT}$)
scale~\cite{KimKyaeKK08}. We take the following forms of a shift
vector and a Wilson line, which are associated with the boundary
conditions of $X_L^I$:
\begin{eqnarray}
&&V=\textstyle\left( \frac14~ \frac14~ \frac14~ \frac{1}{4}~
\frac14~ ; \frac{5}{12}~\frac{5}{12}~  \frac{1}{12}~ \right)\left(
\frac{1}{4}~
\frac{3}{4}~ 0~ 0~0~0~0~0 \right) , \quad\quad  \nonumber \\
&&W=\textstyle\left(
\frac23~\frac23~\frac23~\frac{-2}{3}~\frac{-2}{3}~;\frac23~
0~\frac23 \right)\left( 0~ \frac23~\frac{2}{3}~0~0~0~0~0 \right)
\nonumber \label{Z12ISM} ~.
\end{eqnarray}
They satisfy Eqs.~(\ref{v^2})-(\ref{w^2}). The massless states in
gauge sector are given by the ${\rm E_8\times E_8'}$ root vectors
($P^2=2$) satisfying $P\cdot V={\rm integer}$ and $P\cdot W={\rm
integer}$ with $s\cdot\phi=0$. They are $\textstyle
(\underline{1~-1~0}~;~0~0~;~0^3)(0^8)'$, $\textstyle
(0~0~0~;~\underline{1~-1}~;~0^3)(0^8)'$, $
(0^8)(0~0~;~\underline{\pm 1~\pm 1~0~0~0})'$,  where the
underlined entries allow all possible commutations. Thus, the
resulting gauge group is
\begin{eqnarray}
{\rm [\{SU(3)_c\times SU(2)_L\times U(1)_Y\}\times U(1)^4]\times
[SO(10)\times U(1)^3]^\prime }~. \nonumber
\end{eqnarray}
The hypercharge of ${\rm U(1)_Y}$ is defined with the standard
one, $\textstyle
Y=\sqrt{\frac35}~(\frac13~\frac13~\frac13~\frac{-1}{2}~
\frac{-1}{2};0^3)(0^8)'$. Here the normalization factor
$\sqrt{3/5}$, which leads to ${\rm sin}^2\theta_W=\frac{3}{8}$ at
the string scale, is determined by the current algebra in the
heterotic string theory. The important features of this model are
summarized as follows:
\begin{itemize}
\item
One family of the MSSM matter comes from the $U$ sectors and two
families from $T_4$. They all compose full sets of SO(10) spinor
representations. The other matter form vector-like representations
under $G_{SM}$.
\item
The MSSM two Higgs doublets $(0 0 0~\underline{\pm
1~0};0^3)(0^8)'$ are from the $U$ sector, whereas the triplets
$(\underline{\pm 1~0~0}~0 0;0^3)(0^8)'$ are absent there.
\item
The exact matter parity can be defined on a vacuum, where all
extra U(1)s are broken and all extra states achieve superheavy
masses by SM singlet VEVs.
\item
Above the $1/R$ scale ($\sim M_{\rm GUT}$), the model becomes a 6D
SU(8) SUSY GUT, and the MSSM gauge couplings are unified around
$10^{17}$ GeV. The two Higgs doublets are included in the 6D gauge
multiplet.
\end{itemize}



\bibliographystyle{aipproc}   


\begin{thebibliography}{9}

\def\prp#1#2#3{Phys.\ Rep.\ {\bf #1}}
\def\rmp#1#2#3{Rev. Mod. Phys.\ {\bf #1}, #2 (#3)}
\def\npb#1#2#3{Nucl.\ Phys.\ {\bf B#1}, #2 (#3)}
\def\plb#1#2#3{Phys.\ Lett.\ {\bf B#1}, #2 (#3)}
\def\prd#1#2#3{Phys.\ Rev.\ {\bf D#1}, #2 (#3)}
\def\prl#1#2#3{Phys.\ Rev.\ Lett.\ {\bf #1}, #2 (#3)}
\def\jhep#1#2#3{JHEP\ {\bf #1}, #2 (#3)}
\def\jcap#1#2#3{JCAP\ {\bf #1}, #2 (#3)}
\def\zp#1#2#3{Z.\ Phys.\ {\bf #1}, #2 (#3)}
\def\epjc#1#2#3{Euro. Phys. J.\ {\bf C#1}, #2 (#3)}
\def\ijmp#1#2#3{Int.\ J.\ Mod.\ Phys.\ {\bf #1}, #2 (#3)}
\def\mpl#1#2#3{Mod.\ Phys.\ Lett.\ {\bf A#1}, #2 (#3)}
\def\apj#1#2#3{Astrophys.\ J.\ {\bf #1}, #2 (#3)}
\def\nat#1#2#3{Nature\ {\bf #1}, #2 (#3)}
\def\sjnp#1#2#3{Sov.\ J.\ Nucl.\ Phys.\ {\bf #1}, #2 (#3)}




\bibitem{5dSUSYGUT}
  Y.~Kawamura,
  Prog.\ Theor.\ Phys.\  {\bf 105}, 999 (2001);
L.~J.~Hall and Y.~Nomura,
   \prd{65}{125012}{2002}.
    B.~Kyae, C.~A.~Lee and Q.~Shafi,
  \npb{683}{105}{2004}.

\bibitem{Dixon:1990pc}
  L.~J.~Dixon, V.~Kaplunovsky and J.~Louis,
  \npb{355}{649}{1991};
  K.~S.~Narain, M.~H.~Sarmadi and E.~Witten,
  \npb{279}{369}{1987}.

\bibitem{KimKyaeKK08}
  J.~E.~Kim and B.~Kyae,
  Phys.\ Rev.\  D {\bf 77}, 106008 (2008).

\bibitem{Z12I} J. E. Kim and B. Kyae, \npb{770}{47}{2007}.


\bibitem{SMZ12I} J. E. Kim, J.-H. Kim and B. Kyae,
\jhep{0706}{034}{2007}.
See also J.~E.~Kim, J.~H.~Kim, and B.~Kyae,
 arXiv:0710.4868 [hep-th].

\bibitem{Lebedev}
  O.~Lebedev {\it et al.},
  Phys.\ Rev.\  D {\bf 77}, 046013 (2008).

\end{thebibliography}

\IfFileExists{\jobname.bbl}{}
 {\typeout{}
  \typeout{******************************************}
  \typeout{** Please run "bibtex \jobname" to optain}
  \typeout{** the bibliography and then re-run LaTeX}
  \typeout{** twice to fix the references!}
  \typeout{******************************************}
  \typeout{}
 }


\end{document}